\documentclass{article}

\usepackage{arxiv}

\usepackage[utf8]{inputenc} 
\usepackage[T1]{fontenc}    
\usepackage[colorlinks,urlcolor=black]{hyperref}       
\usepackage{url}            
\usepackage{booktabs}       
\usepackage{amsfonts}       
\usepackage{amsmath}
\usepackage{amssymb}
\usepackage{nicefrac}       
\usepackage{microtype}      
\usepackage{cleveref}       
\usepackage{graphicx}
\usepackage[numbers]{natbib}
\usepackage{doi}
\usepackage{bm}
\usepackage{subfigure}
\usepackage{multirow}
\usepackage{threeparttable}

\title{Weighted Combination and Singular Spectrum Analysis Based Remote Photoplethysmography Pulse Extraction in Low-light Environments}

\date{}

\newif\ifuniqueAffiliation

\ifuniqueAffiliation 
\author{ \href{https://orcid.org/0000-0000-0000-0000}{\includegraphics[scale=0.06]{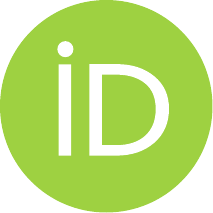}\hspace{1mm}David S.~Hippocampus}\thanks{Use footnote for providing further
		information about author (webpage, alternative
		address)---\emph{not} for acknowledging funding agencies.} \\
	Department of Computer Science\\
	Cranberry-Lemon University\\
	Pittsburgh, PA 15213 \\
	\texttt{hippo@cs.cranberry-lemon.edu} \\
	\And
	\href{https://orcid.org/0000-0000-0000-0000}{\includegraphics[scale=0.06]{orcid.pdf}\hspace{1mm}Elias D.~Striatum} \\
	Department of Electrical Engineering\\
	Mount-Sheikh University\\
	Santa Narimana, Levand \\
	\texttt{stariate@ee.mount-sheikh.edu} \\
}
\else
\usepackage{authblk}

\newbox{\orcid}\sbox{\orcid}{\includegraphics[scale=0.06]{orcid.pdf}} 
\author[1]{%
	\href{https://orcid.org/orcid=0000-0001-6075-5614}{\usebox{\orcid}\hspace{1mm}Lin~Xi}%
}
\author[1]{%
	{Xingming~Wu}%
}
\author[1]{%
	\href{https://orcid.org/orcid=0000-0001-7912-4505}{\usebox{\orcid}\hspace{1mm}Weihai~Chen\thanks{Corresponding authors}}%
}
\author[1]{%
	{Jianhua~Wang}%
}
\author[2]{%
	\href{https://orcid.org/orcid=0000-0002-0546-6016}{\usebox{\orcid}\hspace{1mm}Changchen~Zhao$^{*}$}%
}
\affil[1]{School of Automation Science and Electrical Engineering, Beihang University, Beijing 100191, China \newline \texttt{xilin1991@buaa.edu.cn, xmwubuaa@163.com, whchen@buaa.edu.cn, jhwang@126.com}}
\affil[2]{Hangzhou Innovation Institute, Beihang University, Hangzhou 310051, China \newline \texttt{cczhao@zjut.edu.cn}}
\fi

\renewcommand{\headeright}{}
\renewcommand{\undertitle}{}
\renewcommand{\shorttitle}{}

\hypersetup{
pdftitle={Weighted Combination and Singular Spectrum Analysis Based Remote Photoplethysmography Pulse Extraction in Low-light Environments},
pdfsubject={q-bio.NC, q-bio.QM},
pdfauthor={Lin~Xi, Xingming~Wu, Weihai~Chen, Jianhua~Wang, Changchen~Zhao},
pdfkeywords={Biosignal processing, Remote photoplethysmography, Heart rate estimation, Noise suppression, Low-illumination environments},
}

\begin{document}
\maketitle

\begin{abstract}
	Camera-based vital signs monitoring in recent years has attracted more and more researchers and the results are promising. However, a few research works focus on heart rate extraction under extremely low illumination environments. In this paper, we propose a novel framework for remote heart rate estimation under low-light conditions. This method uses singular spectrum analysis (SSA) to decompose the filtered signal into several reconstructed components. A spectral masking algorithm is utilized to refine the preliminary candidate components on the basis of a reference heart rate. The contributive components are fused into the final pulse signal. To evaluate the performance of our framework in low-light conditions, the proposed approach is tested on a large-scale multi-illumination HR dataset (named MIHR). The test results verify that the proposed method has stronger robustness to low illumination than state-of-the-art methods, effectively improving the signal-to-noise ratio and heart rate estimation precision. We further perform experiments on the PUlse RatE detection (PURE) dataset which is recorded under normal light conditions to demonstrate the generalization of our method. The experiment results show that our method can stably detect pulse rate and achieve comparative results. The proposed method pioneers a new solution to the remote heart rate estimation in low-light conditions.
\end{abstract}


\section{Introduction}\label{sec:Intro}

As one of the most important vital signs of the human body, the heart rate (HR) can reveal a large amount of valuable information, including health status, lifestyle, emotional state, and early onset of heart disease. Typical HR measurements, including electrocardiogram (ECG) \cite{ECG} and photoplethysmography (PPG) \cite{PPG}, have been widely used in many medical care scenarios. Recently, researchers have shown that digital cameras can capture pulse-induced subtle color variations from human skin surfaces \cite{Verkruysse1}. It enables us to measure human cardiac activities in a non-contact way by means of simple instrumentations such as consumer-level digital cameras \cite{Verkruysse1, Poh1, Poh2} and cellphone \cite{Pelegris}. This contactless technology is usually called remote photoplethysmography (rPPG). Since rPPG does not need any direct physical contact with the subject, it can be used for numerous healthcare applications such as driver monitoring \cite{Driver_monitoring}, fitness monitoring \cite{Lin2018StepCA,Wang2016QualityMF,Wang2017RobustHR}, home healthcare \cite{HomeHealth}, and face anti-spoofing \cite{anti1,anti2, anti3}.

Over the last decades, numerous methods have been proposed to extract rPPG signals from video. These methods can be roughly grouped into two main categories: 1) conventional methods, and 2) deep learning-based methods.

A group of conventional methods is based on signal separation techniques to separate pulse signals from observations. The three-channeled signals can be regarded as three mixed sources and the signal separation model can be utilized for pulse extraction, \emph{e.g.}, independent component analysis (ICA) \cite{Poh1}, principal component analysis (PCA) \cite{2011Measuring}, ensemble empirical mode decomposition (EEMD) \cite{2015Image}, normalized least mean square (NLMS) adaptive filter \cite{li2014remote}, or nonlinear mode decomposition (NMD) \cite{Demirezen2021HeartRE}. Another line of work focuses on modeling the image process of photoplethysmography based on physiological and optical principles, namely, the skin reflection model. Based on the model, researchers proposed CHROM \cite{Haan2} and POS \cite{Wang_POS}. Furthermore, there exists one more line of work that is based on the blood volume pulse signature \cite{de_Haan_2014,2018Single}, which represents intrinsic relative pulsatile amplitude in three color channels.

Recently, neural networks have been introduced to rPPG and showed excellent performance. Chen \emph{et al.} \cite{chen2018deepphys} proposed a two-path neural network DeepPhys, which consists of an appearance path for skin localization and a motion path for signal extraction. Huang \emph{et al.} \cite{Huang2020Anovel} proposed PRNet, a one-stage spatio-temporal framework where LSTM modules are used to extract spatial and latent temporal information that is hidden in a video. Nowara \emph{et al.} \cite{Nowara_ICCVW} trained the deep learning models on compressed videos to perform significant experiments that study the influence of parameters of the quality and resolution of videos.

Although the accuracy of previous remote HR monitoring methods has been remarkably improved, those methods have not considered the influence of illumination on rPPG and have poor performance in low-light conditions. To the best of our knowledge, there is a limited number of researches that focus on pulse extraction in low-light conditions using visible light cameras (only \cite{HighSensitive} and \cite{Lin_FG2020}). However, extracting pulse signals under low-light conditions has a wide range of applications and the problem is challenging. Therefore, this problem needs further and in-depth investigation.

Usually, measuring HR in low-light environment typically does not have sufficient or even no light, which makes the rPPG monitoring a challenging task. We are faced with two challenges in low-light environment, which were not observed in sufficient illumination scenarios. The first one is that the rPPG signal captured by the camera is extremely subtle. The amplitude of pulsatile signal has significantly decreased, thereby causing the noise signal to be in a dominant position. Second, the pulse extraction models may fail because their assumption of the blood volume pulse signature no longer holds in low-light situations. The skin reflection model \cite{Wang_POS,Haan2} relies heavily on skin tone direction and pulsatile color variation direction in RGB space. The directions change in low-light conditions because the signals in blue and red channels are approximately zeros. In ICA \cite{Poh1} and PCA \cite{2011Measuring}, the three color channels (RGB) are linearly combined into a pulse signal, whereas in low-light conditions, the red and blue channels contain mainly noise rather than the pulse signal. The existing rPPG methods cannot identify the pulse signal perfectly in the presence of significant noise. Okata \emph{et al.} \cite{HighSensitive} proposed a method of measuring a pulse wave using an ultra-high sensitivity camera in a low illumination environment. To overcome the low illumination, they used an ultra-high sensitivity camera, which put higher requirements on the sensing equipment. Okata \emph{et al.}'s approach is essentially ICA, which does not consider the nature of traces under low-light environment. The previous work \cite{Lin_FG2020} demonstrates that the existing rPPG algorithms are ineffective on low-light videos, which implies that camera quantization noise in low-light conditions poses a serious threat to the pulse signals.

To address the above-mentioned challenges, we propose a novel framework to extract rPPG information in low-light conditions by exploring the frequency characteristics (structure) of rPPG signal and camera quantization noise. This idea originates from the observation that the frequency structures of the rPPG and noise signal remain unchanged even in low-light conditions. In our method, we use an ordinary digital camera that is widely used in most rPPG systems, whereas the ultrahigh sensitivity camera in \cite{HighSensitive} is effective when illuminance is greater than $10^{1.0}\ (=10.0)$ lux, where the scene contents are still visible in the image. However, we investigate the illuminance as low as $10^{0.0}\ (=1.0)$ lux, where the image is almost fully dark. Our solution is necessary for applications such as driver monitoring in the evening, sleep monitoring, and long-term healthcare. These scenarios typically do not have sufficient light for remote HR measurement.

The main contributions of this paper are summarized as follows:
\begin{itemize}
	\item We investigate the influence of light illumination on the rPPG measurement in a mathematical context with optical and physiological reasoning. We provide some insights into the interactions between the noise and pulse signal measured in low-light conditions.
	\item Based on a deep understanding of noise characteristics, we propose the core algorithm for robust rPPG measurement under low-light environment, which separates the rPPG signal in the frequency domain by incorporating a spectral masking algorithm to eliminate the wide-frequency covering noise and removes the noise distributed around the HR by a weighted combination process.
	\item Extensive experiments are conducted on a low-light dataset (large-scale multi-illumination HR dataset, MIHR) and a normal-light dataset (PUlse RatE detection dataset, PURE), which demonstrates the effectiveness of the proposed framework as well as indicates the failure of existing rPPG approaches.
\end{itemize}

The remainder of this article is organized as follows: Section \ref{sec:MET} briefs materials and methods which include an analysis of the camera quantization noise in low-light conditions and an outline of the present framework. Section \ref{sec:EXP} reports the results and discussion. Section \ref{sec:CON} concludes the whole article.

\section{Materials and methods}\label{sec:MET}

\section{Materials and methods}
\label{sec:MET}

\subsection{Analysis of the camera quantization noise}
\label{sec:anal}
Camera quantization noise appears as random fluctuations above and below the actual image intensity. For remote HR estimation, this wide spectral random noise severely affects the quality of the predicted pulse signal. The influence of camera quantization noise is analyzed by comparing the differences in the raw measurement traces under 11 different illuminations. We can observe two distinct characteristics: a) wide frequency coverage and b) equivalent amplitude.

\subsubsection{Wide frequency coverage}
\begin{figure*}[htbp]
	\centering{
		\subfigure[]{
			\includegraphics[width=0.31\linewidth]{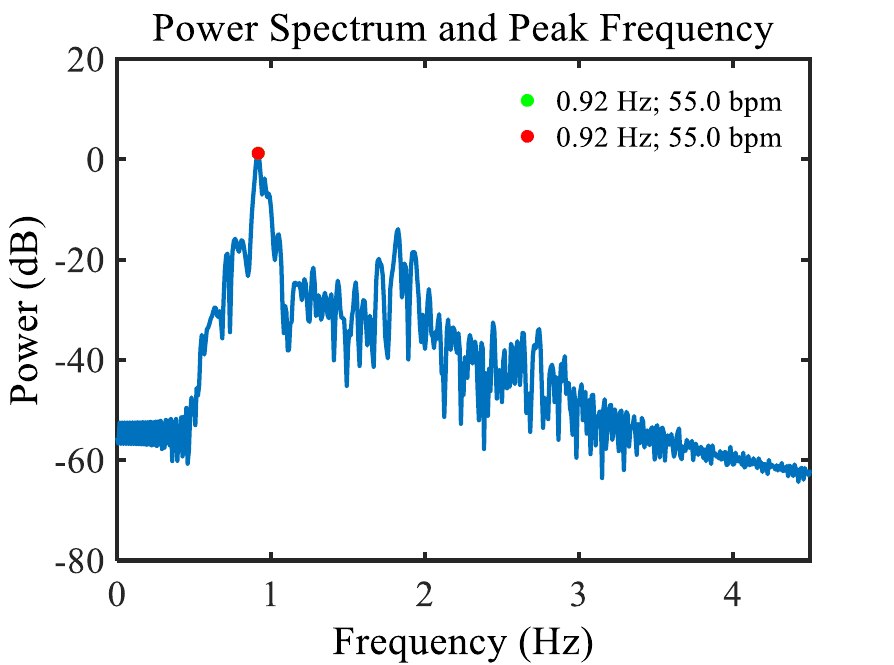}
			\label{fig:widea}
		}
		\subfigure[]{
			\includegraphics[width=0.31\linewidth]{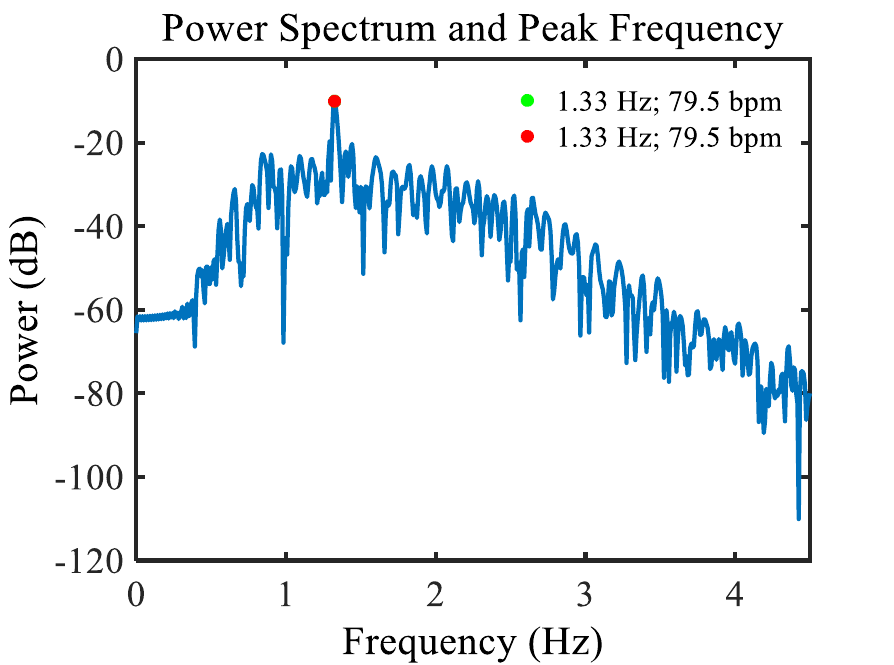}
			\label{fig:widec}
		}
		\subfigure[]{
			\includegraphics[width=0.31\linewidth]{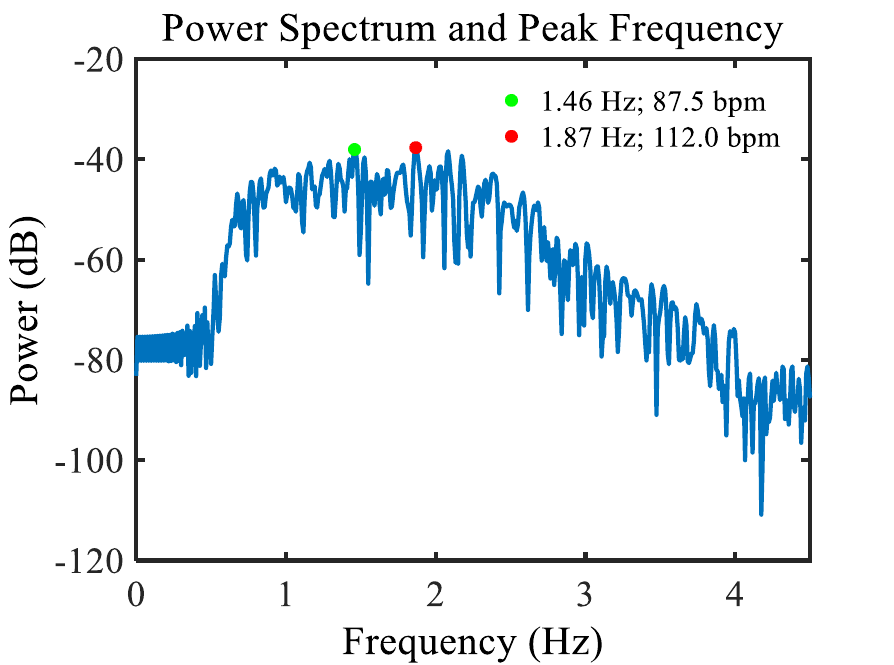}
			\label{fig:widee}
		}
		\quad
		\subfigure[]{
			\includegraphics[width=0.31\linewidth]{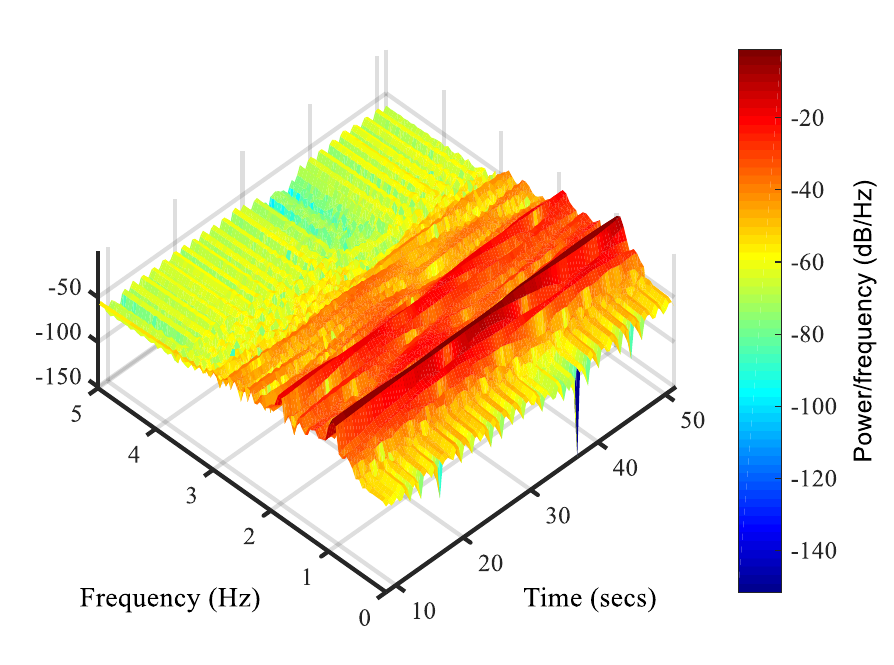}
			\label{fig:wideb}
		}
		\subfigure[]{
			\includegraphics[width=0.31\linewidth]{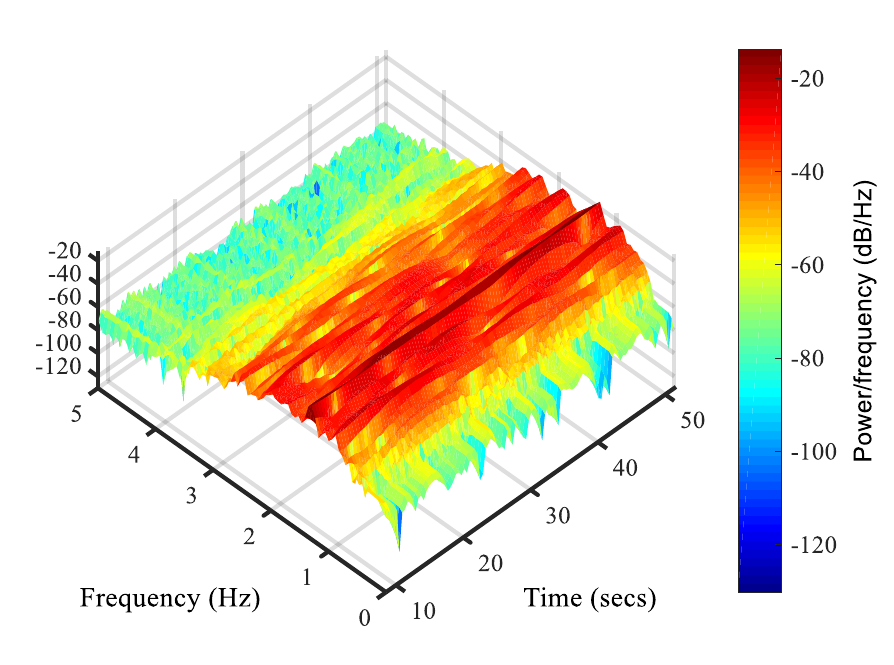}
			\label{fig:wided}
		}
		\subfigure[]{
			\includegraphics[width=0.31\linewidth]{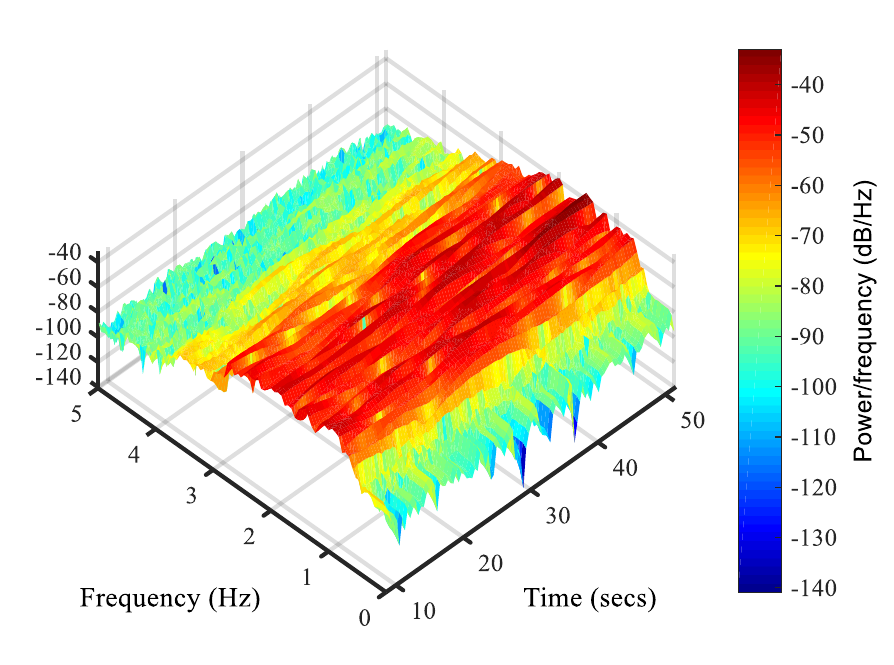}
			\label{fig:widef}
		}
	}
	\caption{The spectrum of raw traces extracted from different illumination conditions. Left column: the frequency spectrum of raw trace; right column: spectrograms of raw trace. First row: illuminance $= 10^{2.0}\ (=100.0)$ lux; second row: illuminance $= 10^{1.0}\ (=10.0)$ lux; last row: illuminance $= 10^{0.0}\ (=1.0)$ lux.}
	\label{fig:wide}
\end{figure*}

The camera quantization noise exhibits a wide frequency coverage, as can be seen in Fig. \ref{fig:wide}, where we plot the spectrum and spectrogram of the raw trace (only green channel) under different illuminance. We can see from the figure that the pulse signal is corrupted by the noise from 0.5 Hz to 4.5 Hz in the frequency domain.

Identifying the frequency components of the pulse signal is increasingly difficult by looking at the spectrums of green channel signal with the decrease in illuminance, \emph{e.g.}, from Fig. \ref{fig:wide} (a) to (b) to (c). The spectrum experiences a serious erosion from $10^{2.0}$ (=100.0) lux down to $10^{0.0}$  (=1.0) lux. The HR estimation is incorrect due to the interference of the noise. As can be seen in Figs. \ref{fig:wide} (d), (e), and (f), the spectrum peak associated with pulse frequency becomes weaker as the illuminance decreases. The pulse peak is buried in the wide-band noise with the decrease in illuminance.

\subsubsection{Equivalent amplitude}
\begin{figure*}[htbp]
	\centering
	\subfigure[]{
		\includegraphics[width=0.31\linewidth]{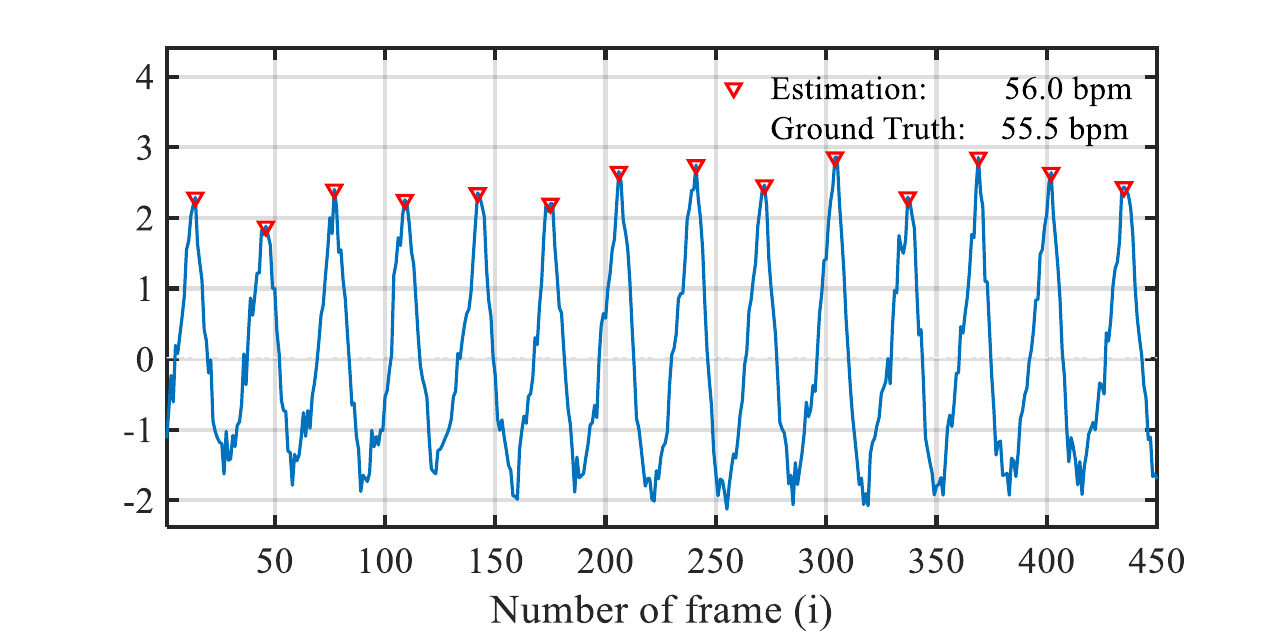}
		\label{fig2:a}
	}
	\subfigure[]{
		\includegraphics[width=0.31\linewidth]{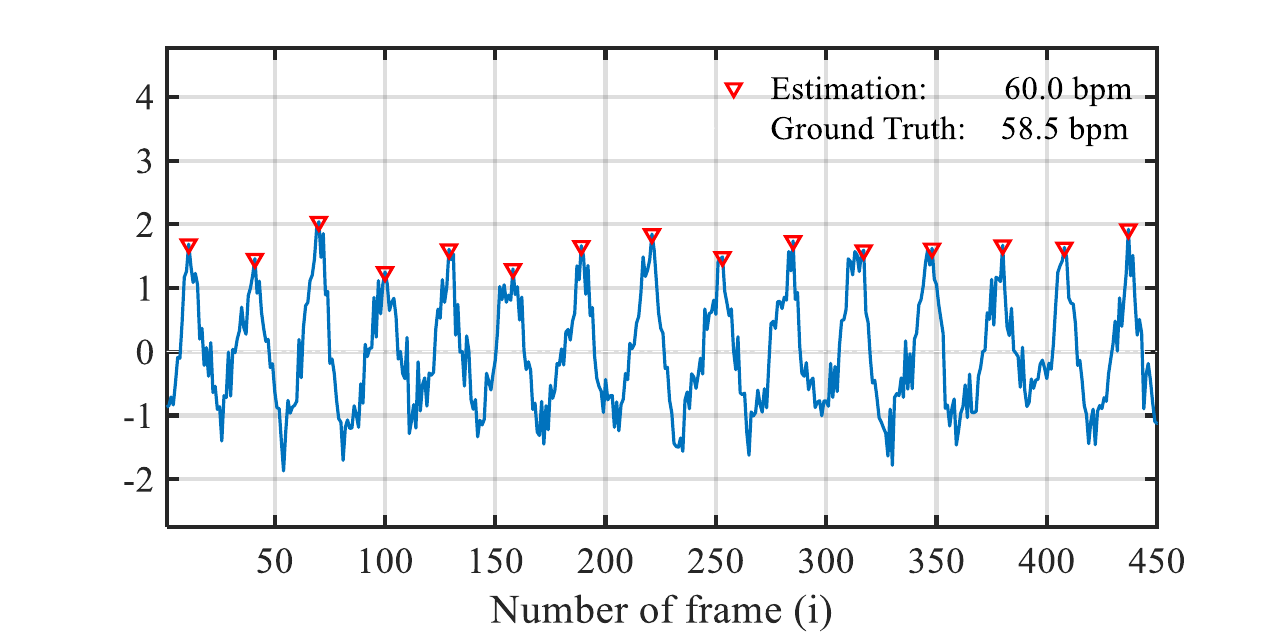}
		\label{fig2:c}
	}
	\subfigure[]{
		\includegraphics[width=0.31\linewidth]{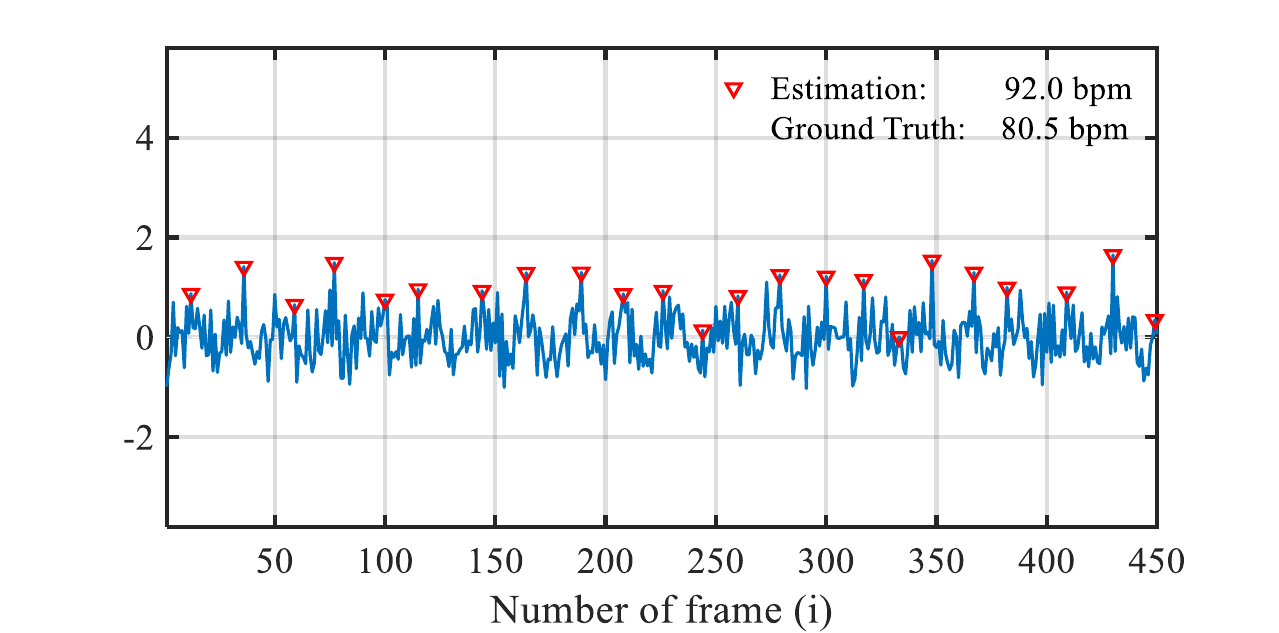}
		\label{fig2:e}
	}
	\quad
	\subfigure[]{
		\includegraphics[width=0.31\linewidth]{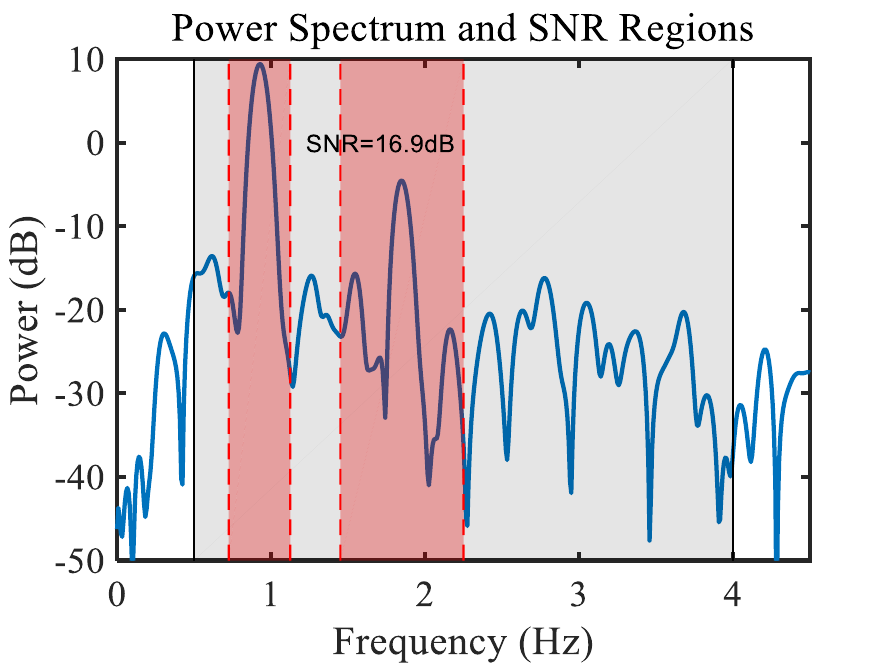}
		\label{fig2:b}
	}
	\subfigure[]{
		\includegraphics[width=0.31\linewidth]{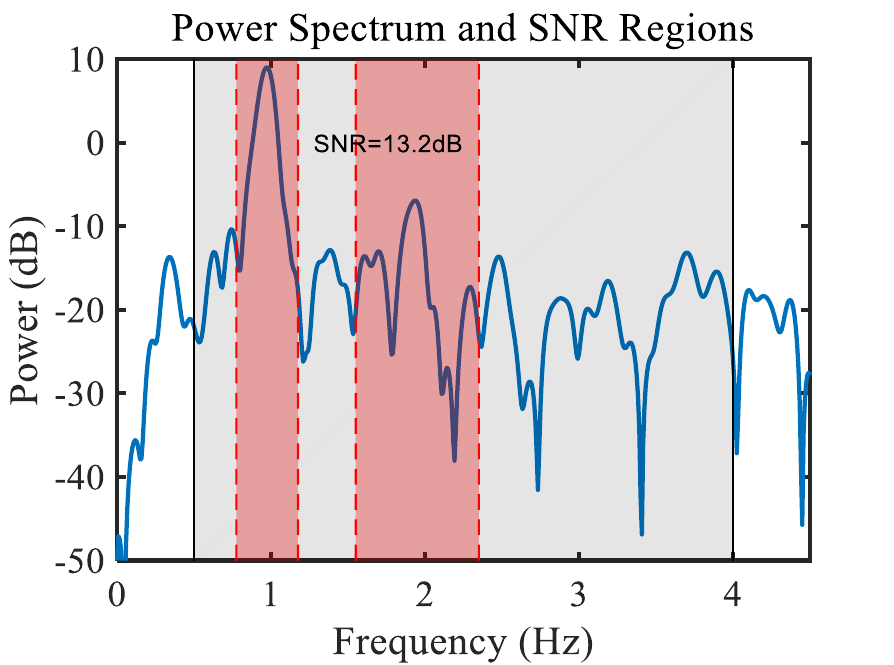}
		\label{fig2:d}
	}	
	\subfigure[]{
		\includegraphics[width=0.31\linewidth]{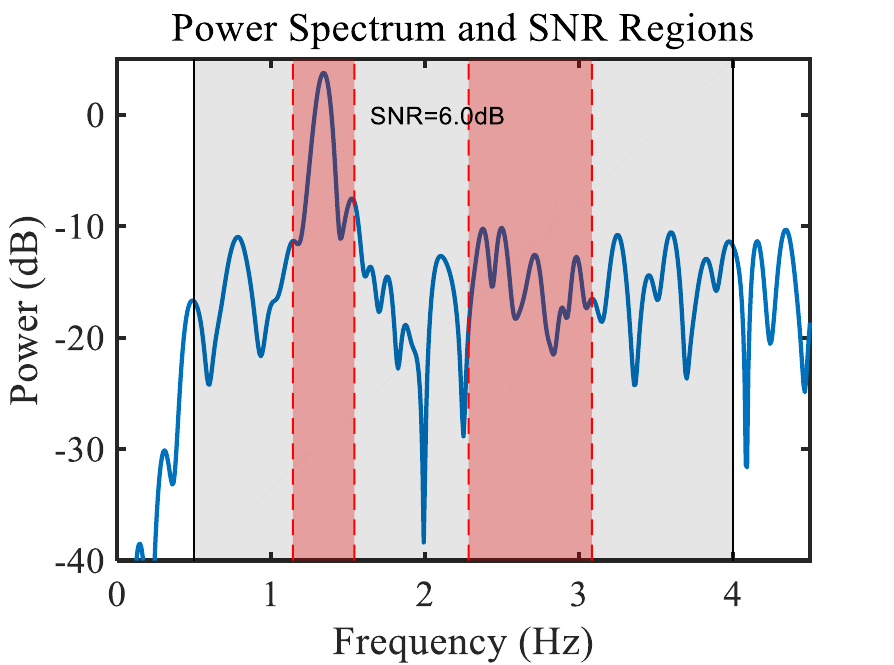}
		\label{fig2:f}
	}
	\caption{Pulse amplitude decreases to be equivalent to that of the noise with the decrease of illuminance. Left column: green channel trace in the time domain. The red triangle denotes the peak output by a peak detection algorithm. Right column: spectrum and SNR region of the signal in the same row. Red-shaded regions denote the ground truth HR frequency and its first harmonic while gray-shaded regions denote the frequency range [0.7, 4] Hz. The SNR is calculated according to the ratio of the spectrum power in the red-shaded region to the residual power in the gray shaded region. First row, illuminance $= 10^{2.0}\ (=100.0)$ lux; second row: illuminance $= 10^{1.4}\ (\approx25.1)$ lux; last row: illuminance $= 10^{0.6}\ (\approx4.0)$ lux.}
	\label{fig:equiv}
\end{figure*}

The pulse amplitude tends to be attenuated which is equivalent to that of the noise with the decrease in illuminance. We can see in Fig.s \ref{fig:equiv} (a), (b), and (c) that a clear deterioration of pulse amplitude is observed with the decrease in illuminance. When the illuminance decreases to $10^{0.6}$ ($\approx$4.0) lux, the amplitudes of pulse and noise are almost the same so that the peak detection algorithm cannot detect the correct pulse peak. As a result, the estimated HR error increases dramatically.

Taking from the spectral domain, \emph{e.g.}, Fig.s \ref{fig:equiv} (d), (e), and (f), the SNR decreases as the illuminance decreases. We can still identify the dominant pulse peak and its first harmonic in Fig.s \ref{fig:equiv} (d) and (e), while the dominant peak in Fig. \ref{fig:equiv} (f) is no longer associated with the pulse and the first harmonic is invisible and buried in the noise power.

\subsection{Proposed Methodology}

\begin{figure*}
	\begin{center}
		\includegraphics[width=1.0\textwidth]{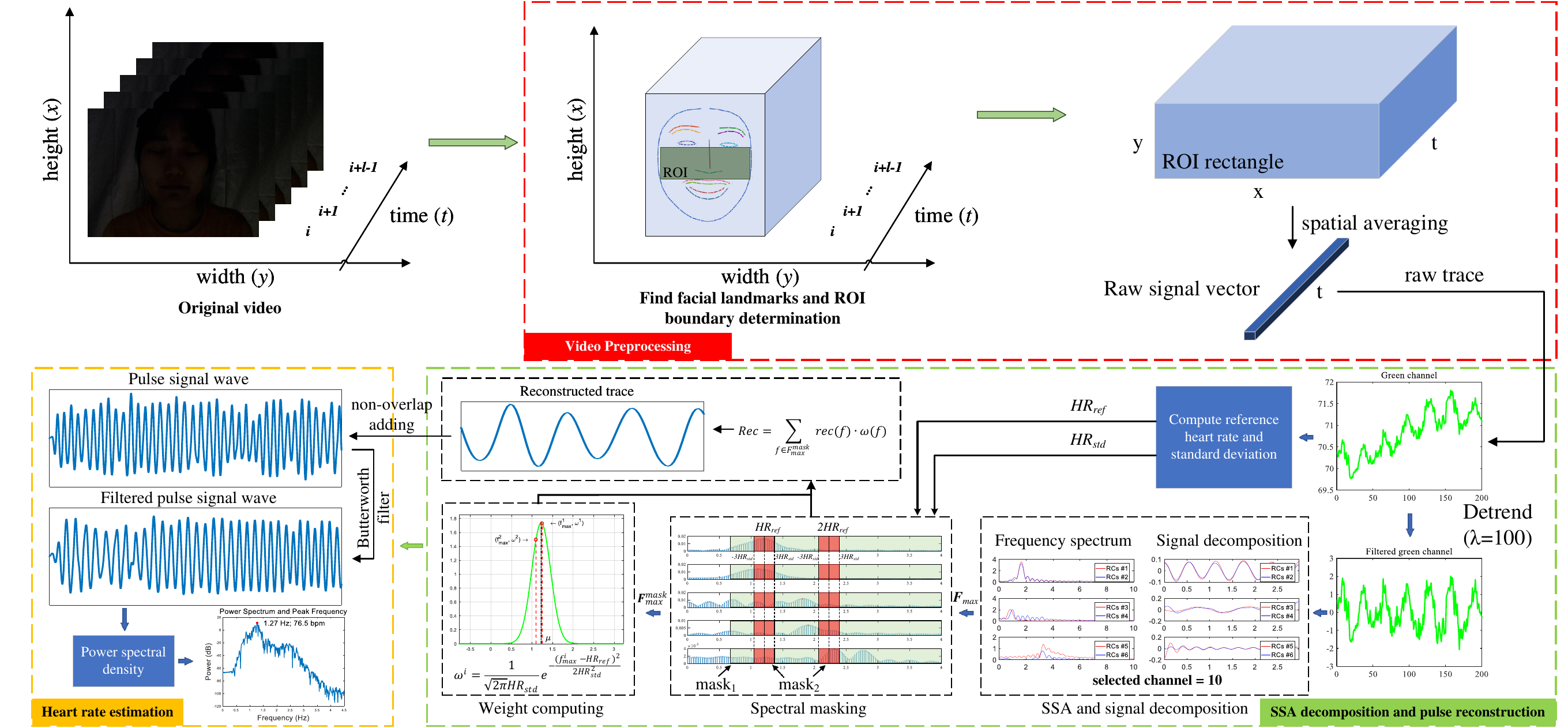}
	\end{center}
	\caption{Flowchart of the proposed framework for pulse extraction under low-light conditions. Given an input video sequence, we first enhance each frame and perform face detection and tracking. The ROI and raw traces are then obtained by spatial averaging. Subsequently, the detrend green channel trace is decomposed using singular spectrum analysis (SSA) and a weighted frequency spectrum of candidate fusion strategy is used to reconstruct the pulse signal. The HR can finally be estimated from the fused pulse signal wave.}
	\label{fig:flowchart}
\end{figure*}

Based on the previous analysis, we come to the following observations. First, the attenuation of the pulse amplitude in three color channels is different, the green channel is the most reliable. Second, a decrease in the amplitude of the pulsatile signal causes the noise signal to be in a dominant position in the red and blue channels. Third, the camera quantization noise exhibits a wide frequency coverage, but it will not destroy the frequency structure of rPPG signal, that is, the second harmonic of the dominant frequency retains the pulse signal and is the valid signal. We have to cater for signal separation methods for pulse extraction. We proposed remote HR estimation method based on signal decomposition and reconstruction, the flowchart of which is shown in Fig. \ref{fig:flowchart}. The framework consists of three major phases: 1) signal decomposition and pulse component selection, 2) pulse weighted reconstruction, and 3) HR estimation. 

Before the HR estimation, the facial skin area is tracked and the raw measurement trace is obtained. Following the widely used protocol \cite{Wang_POS, Lin_FG2020, Zhao_Com}, we apply the face detection algorithm \cite{Viola_Jones1, Viola_Jones2, Asthana} to generate a rectangular mask of the region of interest (ROI). A spatial averaging \cite{Wang_POS} is then used to obtain a vector resulting in a column vector $\bm{c}(i) \in \mathbb{R}^{3}$, where $i$ and $3$ denote the index of the frame and the color channel, respectively. The raw traces $\bm{C} \in \mathbb{R}^{3\times T}$ are obtained by concatenating all the vectors of each frame.
\begin{equation}
	\bm{C}=\left [ \bm{c}(1), \bm{c}(2), ..., \bm{c}(T) \right ]
	\label{equ:C}
\end{equation}
where $T$ denotes the number of frames in a time window. Subsequently, the raw signal $\bm{C}$ is detrended using the method in \cite{DetrendedMet} (with $\lambda$=100). The detrended trace can be denoted as the following equation:
\begin{equation}
	\widetilde{\bm{C}}=\left [ \widetilde{\bm{c}}(1), \widetilde{\bm{c}}(2), ..., \widetilde{\bm{c}}(T) \right ]
	\label{equ:C_tilda}
\end{equation}

\subsubsection{SSA decomposition and pulse component selection}
Due to the fact that the traces in blue and red channels have large noise, we only use the green channel trace for the subsequent processing, and simply denote it as $\bm{x}=\widetilde{\bm{c}}_g$.  The purpose of SSA signal decomposition is to decompose the detrended trace $\bm{x} = [x_{1},...,x_{T}]$ into a sum of $S$ components. Specifically, SSA maps the detrended trace into a $L \times K$ matrix $\bm{X}$, \emph{i.e.}, Hankel matrix embedding, where $K = T - L + 1$, $L < T/2$, called L-trajectory matrix \cite{SSABook}. The L-trajectory matrix is decomposed by singular value decomposition (SVD) as follows:
\begin{equation}
	\bm{X}=\sum_{i=1}^{r}{\bm{X}_{i}},\ \ \ r=min\{L,K\},
\end{equation}
where $\bm{X}_{i}=\sigma_{i}\bm{u}_{i}\bm{v}_{i}^{T}$, and $\sigma_{i}$, $\bm{u}_{i}$, $\bm{v}_{i}$ denote the $i$-th singular value of $\bm{X}$ and the left and right singular vectors, respectively. The $r$ matrix $X_{i}$, where matrix $X_{i}$ has rank $1$, is assigned to $g$ group. Specifically, the set of indices $\{1,...,r\}$ is partitioned into $g$ disjoint subsets $\{I_{1},...,I_{g}\} (g \leq r)$ and
\begin{equation}
	\bm{X} = \sum_{p=1}^{g}{\bm{X}_{I_{p}}},
\end{equation}
with $\bm{X}_{I_{p}}=\sum_{t \in I_{p}}{\bm{X}_{t}}$. The rank-one matrices in each group $\bm{X}_{I_{p}}$ satisfy some common characteristics. The collection components after reconstruction have the same frequency or exhibit harmonic relation. In the reconstruction step, each group $\bm{X}_{I_{p}}$ is used to reconstructing a time series $\widetilde{\bm{x}}_{p}$ with length $T$ by using a so-called diagonal averaging procedure \cite{SSABook}. Thus, the original signal $\bm{x}$ can be denoted by the sum of $g$ time series,
\begin{equation}
	\bm{x}=\sum_{p=1}^{g}{\widetilde{\bm{x}}_{p}}.
\end{equation}
We define noise time series $\{\widetilde{\bm{x}}_{p}, p \in I_{noise}\}$, where $I_{noise}$ denotes the set of indices of all noise time series. Consequently, we can rebuild a noise-free signal with the components preserving the most information of the pulsatile, that is,
\begin{equation}
	\bm{x}_{pulse} = \sum_{p \notin I_{noise}}{\widetilde{\bm{x}}_{p}}.
\end{equation}

SSA reconstructs pulsatile components from 0.7 Hz to 4 Hz, and the noise components with frequencies outside of this range are removed with a third-order Butterworth bandpass filter. To select the pulsatile components in the decomposition, we exploit the reference HR and standard deviation of the instantaneous HR computed from the green channel signal to denoise the artifact components. Details are given below.

The candidate components are sorted in accordance with corresponding singular values in descending order. The first 10 ($secChn=10$) components are selected because they preserve the most information of the input signal. The main reason for selecting the first 10 components is that it remains more than $90\%$ of the energy, and the components after the 10th do not retain the pulse signal. The reference HR and the rPPG signal frequency structure are utilized as the criteria for the selection of components. In low-light remote HR estimation, directly using the spectral masking algorithm in \cite{Zhao_Com} cannot fully exploit the frequency structure of the rPPG signal. Thus, we modify the window definition and make the following changes.

Note that some selected dominant frequencies may be close to the dominant and first harmonic frequencies of the reference HR. Thus, the masking procedure may remove the candidate components associated with the HR. The rPPG signal frequency structure characteristics needed to be considered. The first harmonic frequency is selected as twice the dominant frequency. The candidate component pairs where their dominant frequencies have the twice relationship are selected, whereas others are rejected. If the above criterion is not satisfied, the first harmonic frequencies are retained in case of harmonic-related information loss. In the current time window, the instantaneous HR is used to determine the dominant frequencies of signals related to the pulse signal. To ensure the calculation of instantaneous reference HR and sufficient window length for fast Fourier transform (FFT), a sliding window of length $l=10 s$ and step size $1 s$ are applied to locate a windowed trace. The dominant frequency was computed by transforming the windowed trace via FFT to the frequency domain. We set the instantaneous reference HR as the central frequency of spectral masking and updated it every $1 s$. Let the reference HR be $f_{r}$ and the set of candidate components frequencies be $F=\{f_{i}\}_{i=1}^{secChn}$. The spectral masking index function is expressed as follows:
\begin{equation}
	\bm{1}_{f_i}=
	\left\{\begin{matrix}
		1, &
		\begin{matrix}
			f_{r}-3\cdot\sigma(f_{r}) \leq f_{i} \leq f_{r}+3\cdot\sigma(f_{r}) \\
			f_{r}-3\cdot\sigma(f_{r}) \leq \frac{f_{i}}{2} \leq f_{r}+3\cdot\sigma(f_{r}) \\
			0.7 \leq f_{i} \leq  4
		\end{matrix} \\
		0, & \rm{otherwise}
	\end{matrix}\right.,
\end{equation}
where $\sigma(f_{r})$ denotes the standard deviation of instantaneous HR, which is defined as the window length of the spectral mask. In the first window, $\sigma(f_{r})$ is initialized to $0.05$. The candidate components are selected by this indexing function for reconstructing the output signal. The final pulse waveform is then obtained by weighted reconstruction and overlap addition.

\subsubsection{Weighted reconstruction}
\label{subsec:3_4}
To precisely estimate HR, we modulate the reconstruction procedure by using a weight function generated by the Gaussian distribution estimated by the reference HR. We make an assumption that the variation in the instantaneous HR is subject to Gaussian distribution in a short period, which is expressed as follows:
\begin{equation}
	{\rm{HR}} \sim N(\mu_{{\rm{HR}}_{r}}, \sigma_{{\rm{HR}}_{r}}),
\end{equation}
where $\mu_{{\rm{HR}}_{r}}$, $\sigma_{{\rm{HR}}_{r}}$ denote the reference HR and standard deviation, respectively. We can filter the large fluctuation in the frequency of candidate components by generating a weighting function $w(f)$ for the pulse reconstruction. The weights for each component can be determined based on the idea of Gaussian distribution. If the candidate components have a large deviation from reference HR, they are unlikely to be pulse-related and small weight should be assigned. If the component is close to the reference HR, it is likely to be the correct one and a large weight should be assigned. The weighted function $w(f)$ is generated in accordance with the above-mentioned distribution,
\begin{equation}
	w(f) = \frac{1}{\sqrt{2\pi}\sigma_{{\rm{HR}}_{r}}}\exp{\left(-\frac{(f-\mu_{{\rm{HR}}_{r}})^{2}}{2{\sigma_{{\rm{HR}}_{r}}}^{2}}\right)}.
\end{equation}

For every windowed trace, we can combine the selected components to a time series via weighted averaging. With the reference HR as the weight, the pulse trace can be represented as the weighted sum of these corresponding components, which is,
\begin{equation}
	\bm{x}_{w} = \frac{\sum_{p \notin I_{noise}}{w(f_{p})} \odot \widetilde{\bm{x}}_{p}}{\sum_{p \notin I_{noise}}{w(f_{p})}},
\end{equation}
where $\odot$ denotes the elements-wise multiplication. The windowed trace $\bm{x}_{w}$ is concatenated to the final pulse wave $\bm{x}_{pulse}$. Every windowed trace $\bm{x}_{w}$ is scaled using a Hanning window of the same length as the windowed trace, and the overlap adding step is half of the window length.

\subsubsection{HR estimation}
\label{subsec:3_5}

The HR can be estimated from the frequency spectrum of the final pulse wave. We use FFT to compute the frequency spectrum and the frequency with the highest power is regarded as the HR. And then, it is transformed into the unit of beats per minute (bpm),
\begin{equation}
	\begin{matrix}
		{\rm{HR}} = 60 \times\mathop{\arg\max}\limits_{f}{\bm{p}(f)} \\
		s.t.\ \emph{\emph{0.7}} \leq f \leq  \emph{\emph{4.0}}
	\end{matrix}
\end{equation}
where $\bm{p}(f)$ denotes the Fourier transform of pulse wave $\bm{x}_{pulse}$.

\section{Results and discussion}
\label{sec:EXP}
This section presents the results of our proposed method compared with the existing methods in the public benchmarks of MIHR \cite{Lin_FG2020} and PURE \cite{PURE}. Then, we discuss the influence of camera parameters on the proposed algorithm. Three common metrics, such as SNR \cite{Haan2}, and precision (\emph{i.e.,} mean absolute error and root mean squared error), are used in this paper to evaluate the performance of remote HR estimation.

\subsection{Compare with existing methods}
\label{sebsec:5_3}
We first perform a comparative experiment by using 11-level illumination videos in the MIHR dataset and compare the proposed method with three state-of-the-art methods (Green \cite{Verkruysse1}, ICA \cite{Poh1}, POS \cite{Wang_POS}) and enhancement methods \cite{HE, LIME, Lin_FG2020}, respectively named HE, LIME, and Lin. We then perform a comparative experiment with static and dynamic scenarios on the PURE dataset.

\subsubsection{Experiments on the MIHR dataset}
\begin{figure}[!htbp]
	\centering{
		\includegraphics[width=0.5\linewidth]{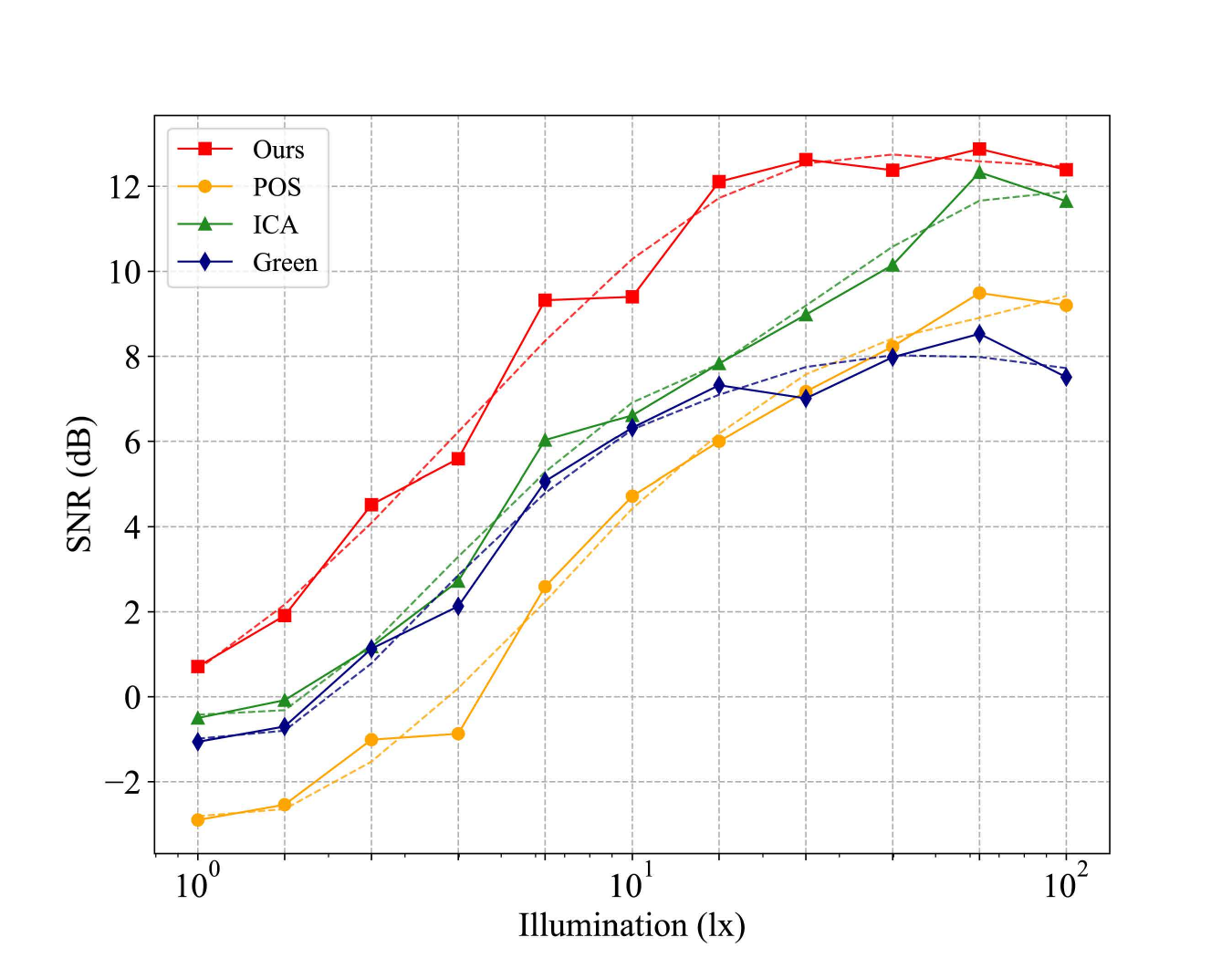}
	}
	\caption{The plot of BVP SNR result as a function of illumination. The \textit{\textbf{solid}} line represents the original data, and the \textit{\textbf{dotted}} line represents the data smoothed by the Savitzky–Golay filter \cite{SAGOFilter}.}
	\label{fig:final}
\end{figure}

We investigate the accuracy and SNR of the proposed method in comparison with state-of-the-art rPPG and enhancement methods under different illumination levels. The pulse signals are extracted from the video using the proposed approach, Green \cite{Verkruysse1}, ICA \cite{Poh1}, POS \cite{Wang_POS}, HE\&Green \cite{HE}, LIME\&Green \cite{LIME}, and Lin\&Green \cite{Lin_FG2020}, respectively. The mean absolute error (MAE), root mean squared error (RMSE), and SNR results as a function of the illuminance are shown in Table \ref{tab:final}. As can be seen, our method outperforms all previous methods by a large margin and achieves a significant improvement in SNR, especially in conditions where the illumination is poor. In order to discover the rPPG algorithm's characteristics from the SNR curve, we plot the SNR result of Green, ICA, POS, and ours as shown in Fig. \ref{fig:final}. Each SNR curve, which follows an 'S'-shape, has a slow growth rate at two ends and a high growth rate in the middle. The main difference is that the turning points are different. The earlier the turning points ($P_{1st}$ and $P_{2nd}$) appear, the better the algorithm is robustness to illumination changes. The two turning points of our method are $(10^{0.0}, 10^{0.8})$ lux, and that of Green, ICA, and POS are $(10^{0.2}, 10^{1.0})$ lux, $(10^{0.2}, 10^{1.8})$ lux, and $(10^{0.2}, 10^{1.4})$ lux, respectively. Our method is more tolerant to the changes in illuminance.

\begin{table*}[htbp]
	\centering
	\footnotesize
	\caption{Summary of the BVP SNR, MAE, and RMSE for the proposed method and several state-of-the-art methods on the MIHR dataset with various light intensity. The best score of each column is highlighted in boldface.}
		\begin{tabular}{ccccccccccccc}
			\toprule
			\multicolumn{2}{c}{\textbf{Illumination} (lux)} & $10^{0.0}$ & $10^{0.2}$ & $10^{0.4}$ & $10^{0.6}$ & $10^{0.8}$ & $10^{1.0}$ & $10^{1.2}$ & $10^{1.4}$ & $10^{1.6}$ & $10^{1.8}$ & $10^{2.0}$   \\ \toprule
			\multirow{7}[0]{*}{\textbf{SNR} (dB) $\uparrow$} & \textit{Green} & -1.06  & -0.75  & 1.13  & 2.13  & 5.06  & 6.32  & 7.32  & 7.01  & 7.98  & 8.53  & 7.51  \\
			& \textit{ICA}   & -0.50  & -0.08  & 1.18  & 2.72  & 6.03  & 6.61  & 7.83  & 8.98  & 10.15  & 12.33  & 11.65  \\
			& \textit{POS}   & -2.90  & -2.54  & -1.01  & -0.87  & 2.59  & 4.72  & 6.00  & 7.17  & 8.23  & 9.49  & 9.20  \\
			& \textit{HE\&Green} & -1.37  & 0.41  & 0.30  & 1.24  & 3.55  & 4.32  & 6.07  & 5.17  & 5.65  & 5.45  & 4.15  \\
			& \textit{LIME\&Green} & -0.19  & 0.03  & 1.00  & 0.11  & 1.45  & 3.72  & 5.38  & 6.61  & 6.63  & 9.46  & 6.70  \\
			& \textit{Lin\&Green} & -0.26  & 1.57  & 1.30  & 1.63  & 3.62  & 3.07  & 3.65  & 3.28  & 3.16  & 2.91  & 4.04  \\
			& \textit{Ours}  & \textbf{0.71}  & \textbf{1.91}  & \textbf{4.52}  & \textbf{5.59}  & \textbf{9.32}  & \textbf{9.40}  & \textbf{12.11}  & \textbf{12.63}  & \textbf{12.38}  & \textbf{12.88}  & \textbf{12.39}  \\ \midrule
			\multirow{7}[0]{*}{\textbf{MAE} (bpm) $\downarrow$} & \textit{Green} & 14.93  & 14.06  & 7.87  & 5.35  & 2.45  & 2.24  & 2.35  & 2.41  & 1.72  & 1.47  & 1.76  \\
			& \textit{ICA}   & 14.96  & 14.54  & 8.04  & 6.76  & 3.59  & 3.39  & 2.32  & 2.34  & 1.80  & 1.17  & 1.69  \\
			& \textit{POS}   & 59.89  & 51.38  & 29.37  & 23.98  & 5.11  & 1.70  & 1.87  & 1.27  & 0.97  & 0.86  & 0.94  \\
			& \textit{HE\&Green} & 18.10  & 12.66  & 8.12  & 8.28  & 4.07  & 3.32  & 2.62  & 2.68  & 3.03  & 2.87  & 3.72  \\
			& \textit{LIME\&Green} & 15.55  & 12.27  & 7.18  & 8.86  & 6.51  & 3.23  & 6.56  & 1.99  & 3.81  & 1.20  & 4.36  \\
			& \textit{Lin\&Green} & 12.12  & 6.01  & 6.30  & 5.35  & 2.97  & 3.06  & 4.48  & 5.21  & 5.60  & 5.73  & 4.94  \\
			& \textit{Ours}  & \textbf{9.47}  & \textbf{5.96}  & \textbf{3.97}  & \textbf{2.32}  & \textbf{0.71}  & \textbf{0.86}  & \textbf{0.56}  & \textbf{0.47}  & \textbf{0.46}  & \textbf{0.45}  & \textbf{0.42}  \\ \midrule
			\multirow{7}[0]{*}{\textbf{RMSE} (bpm) $\downarrow$} & \textit{Green} & 18.98  & 18.32  & 11.91  & 8.71  & 4.13  & 3.10  & 4.20  & 3.88  & 3.09  & 2.89  & 2.63  \\
			& \textit{ICA}   & 20.24  & 19.32  & 12.43  & 8.91  & 5.06  & 4.88  & 3.31  & 3.45  & 2.96  & 1.77  & 2.60  \\
			& \textit{POS}   & 74.88  & 64.73  & 44.53  & 37.02  & 11.36  & 4.01  & 3.88  & 1.71  & 1.34  & 1.17  & 1.19  \\
			& \textit{HE\&Green} & 22.57  & 16.65  & 12.11  & 12.65  & 6.30  & 5.78  & 4.40  & 4.92  & 5.73  & 5.47  & 6.99  \\
			& \textit{LIME\&Green} & 19.69  & 17.10  & 10.95  & 13.32  & 11.28  & 5.33  & 8.09  & 3.08  & 5.77  & 1.81  & 5.88  \\
			& \textit{Lin\&Green} & 17.18  & 10.13  & 9.96  & 9.56  & 4.74  & 5.73  & 7.52  & 7.35  & 8.30  & 9.56  & 8.02  \\
			& \textit{Ours}  & \textbf{11.60}  & \textbf{7.45}  & \textbf{5.16}  & \textbf{3.64}  & \textbf{1.22}  & \textbf{1.56}  & \textbf{1.11}  & \textbf{0.97}  & \textbf{0.89}  & \textbf{0.79}  & \textbf{0.75}  \\ \bottomrule
		\end{tabular}
	\label{tab:final}
\end{table*}

We show the HR estimation accuracy of the proposed method and the compared algorithms in the MAE and RMSE items of Table \ref{tab:final}. The error including MAE and RMSE of all methods goes higher when the illuminance is less than $10^{0.8}$ ($\approx$6.3) lux, suggesting that insufficient lighting limits the performance of HR estimation. Specifically, the MAE and RMSE of POS increase from 5.11 bpm to 59.89 bpm and from 11.36 bpm to 74.88 bpm, respectively. From $10^{0.0}$ (=1.0) lux to $10^{0.8}$ ($\approx$6.3) lux, the precision of our method is the highest, followed by Lin\&Green, ICA, Green, and POS. The experimental results show that the proposed approach can achieve satisfying performance for HR measurement in low-light conditions than the previous methods. The lack of pulsatile signal in the red and blue channels are harmful to the method relying on three color channel combination in the case of illumination less than $10^{0.8}$ ($\approx$6.3) lux. The noise involved in red and blue channels is mixed in the process of pulse extraction. From $10^{1.0}$ (=10.0) lux to $10^{2.0}$ (=100.0) lux, the precision of the proposed method obtains better results and is much better than the compared methods. The proposed method obtains the best results in all lighting cases. All the HR estimation algorithms achieve a slight improvement in precision under the illumination of $10^{1.0}$ (=10.0) lux to $10^{2.0}$ (=100.0) lux. Therefore, our method precisely identifies the pulse components in the low-light condition and provides accurate and stable pulse rate detection. 

\begin{figure}
	\begin{center}
		\includegraphics[width=0.5\linewidth]{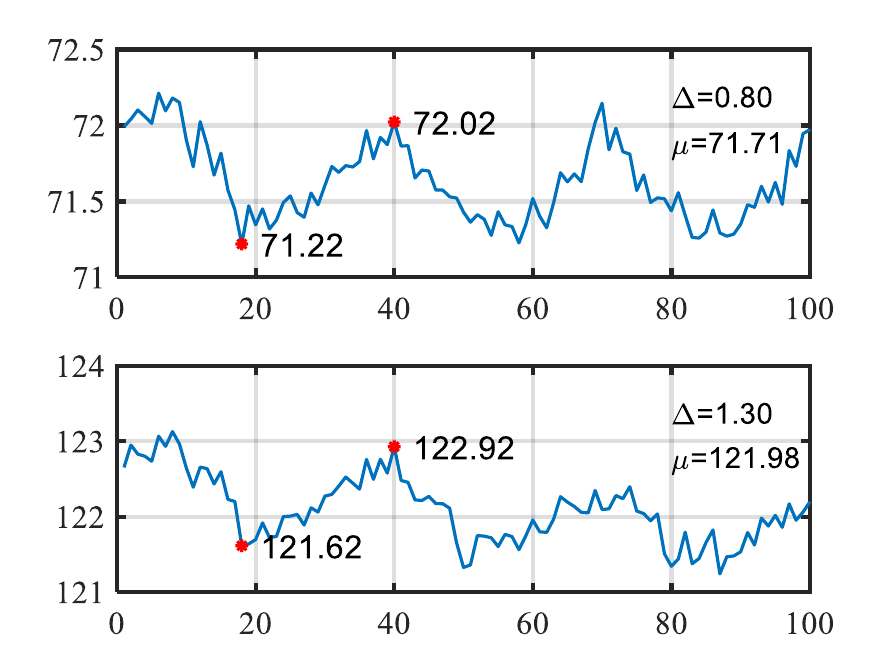}
	\end{center}
	\caption{Raw traces extracted from the same video but processed by different image enhancement methods. First row: original (low-light) video, peak-to-peak value $\Delta_{1}=0.80$, mean of raw trace $\mu_{1}=71.71$; second row: original video enhanced by HE, peak-to-peak value $\Delta_{2}=1.30$, mean of raw trace $\mu_{2}=121.98$. After the HE, we can obtain the ratio of $\mu_{2}$ to $\mu_{1}$ as $\eta_{1}= \frac{\mu_{2}}{\mu_{1}} = 1.70$. The ratio of $\delta_{2}$ to $\delta_{1}$ is $\eta_{2}=\frac{\delta_{2}}{\delta_{1}} = 1.63$. We can obtain the result as $\eta_{1} \approx \eta_{2}$.}
	\label{fig:traceEnhance}
\end{figure}

Compared with the enhancement strategies (HE, LIME, and Lin), our proposed method is also the best. As expected, decreasing the illuminance steadily degrades the video quality, and, in turn, lowers the SNR values. The SNR of the Lin\&Green method is greater than other methods (HE\&Green, LIME\&Green) when the illuminance is less than $10^{0.8}$ ($\approx$6.3) lux. HE\&Green and Lin\&Green improve the accuracy of ROI detection, but only the SNR of the latter increases in low-light conditions. To explain the result that enhancement methods have little effect on pulse extraction, we plot two traces in Fig. \ref{fig:traceEnhance}, where the first row is the trace from the original low-light video and the second row is the trace from the same video but enhanced by HE. We can see that the overall waveform of the two traces is almost the same. The ratio of the peak-to-peak difference and the trace mean are almost the same, implying that the image enhancement method multiplies each video frame by the same factor. This operation has a limited contribution to pulse signal extraction.

According to the experimental results of the MIHR dataset, our proposed method presented strong robustness to the illumination disturbances.

\subsubsection{Experiments on the PURE dataset}

To validate the performance of the proposed method, we test on the PURE dataset in static and dynamic scenarios. We conduct a quantitative statistical comparison and qualitative comparison between the proposed method and state-of-the-art methods to answer the following questions: 1) whether the proposed method can extract HR accurately under normal-light conditions? 2) what is the difference between the camera quantization noise in low-light and the motion artifacts in normal-light conditions?

\begin{figure*}[htbp]
	\centering{
		\subfigure[SNR]{
			\includegraphics[width=0.3\textwidth]{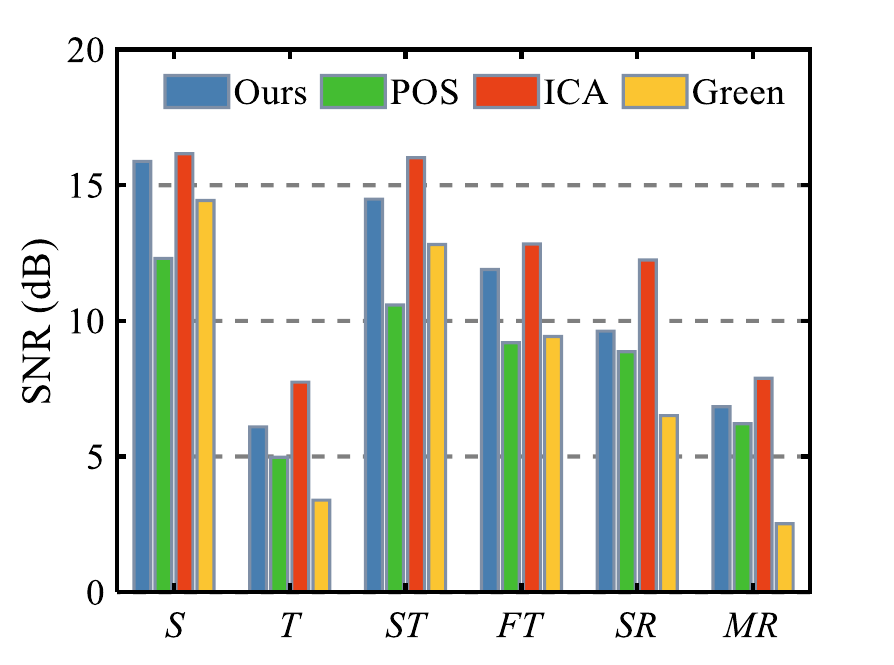}
			\label{fig:PURE_res:a}
		}
		\subfigure[MAE]{
			\includegraphics[width=0.3\textwidth]{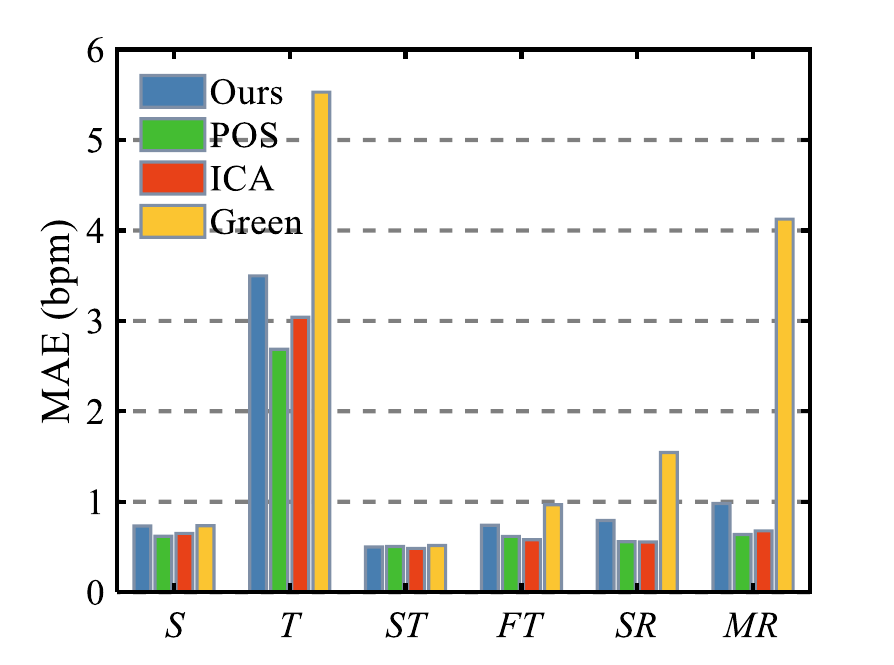}
			\label{fig:PURE_res:b}
		}
		\subfigure[RMSE]{
			\includegraphics[width=0.3\textwidth]{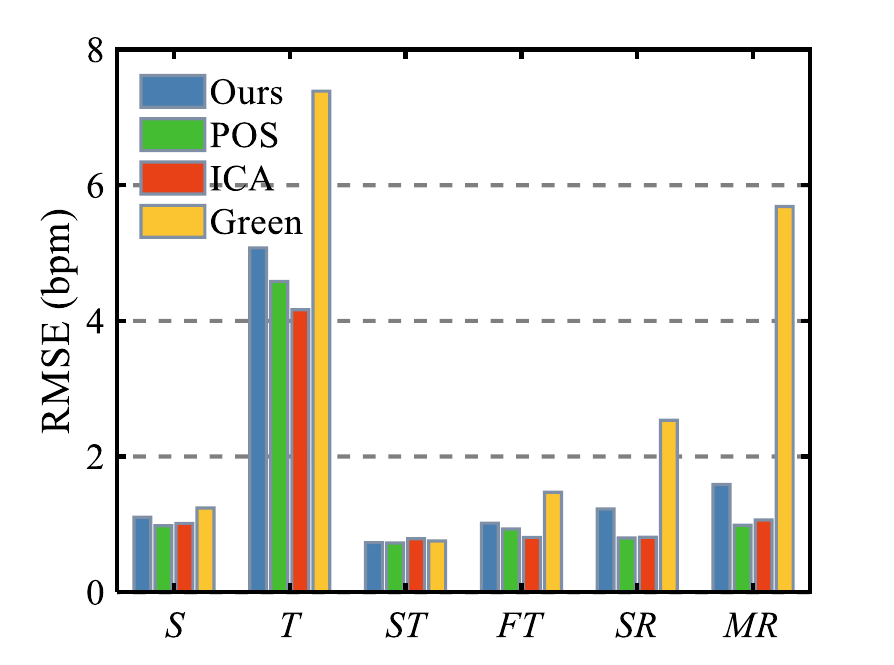}
			\label{fig:PURE_res:c}
		}
	}
	\caption{SNR and precision results of the comparative experiment on the PURE dataset.}
	\label{fig:PURE_res}
\end{figure*}

To answer the first question, we calculate the SNR, precision (MAE and RMSE) of the compared methods. The results are reported in Fig. \ref{fig:PURE_res}. The results show that for the static scenario, POS and ICA perform slightly better than the proposed algorithm in precision, whereas, for the SNR, the proposed method achieves the best result after ICA. In the dynamic scenario, the performance of the proposed algorithm is relatively stable, and the error gap with the best method becomes a small margin in the case of \textit{ST} and \textit{FT}. Similarly, the SNR of our method ranks second in the dynamic cases. The reason is easy to explain. ICA separates the pulse signal based on RGB channels so that a more useful physiological signal can participate in the HR extraction process. However, the proposed method only uses the green channel as the input signal for HR estimation. Even in this case, the SNR is higher than POS, which is based on the three-channel combination. For steady or slow movement cases, the noise caused by illuminance does not affect the HR estimation. Although our method cannot highlight its advantages, it shows high accuracy and stability in pulse rate detection on normal light video. Hence, we can conclude the results that the proposed algorithm can precisely identify the pulse signal from raw trace and provide an excellent quality physiological signal.

\begin{figure}[!htb]
	\begin{center}
		\includegraphics[width=0.5\linewidth]{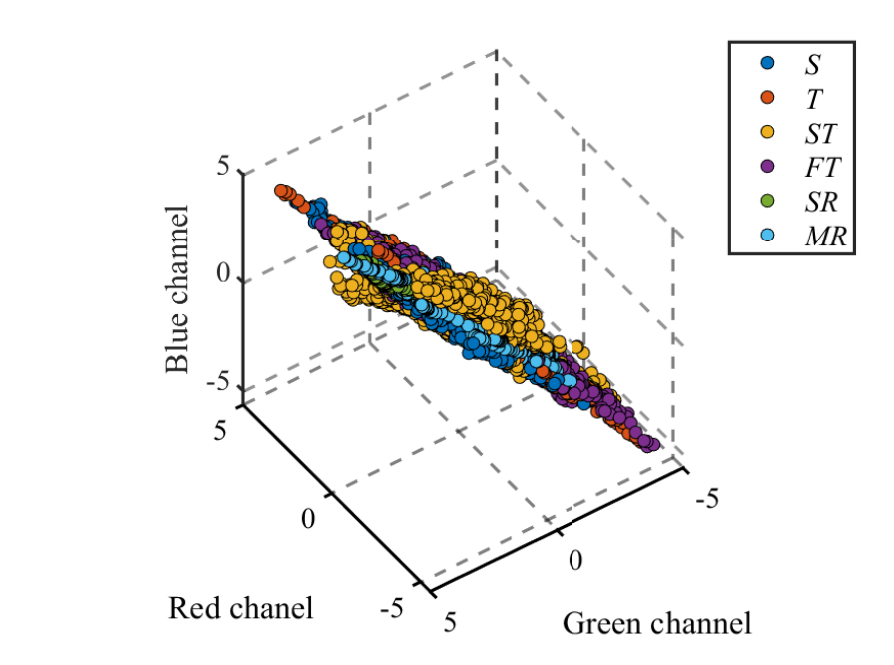}
	\end{center}
	\caption{Detrended trace layout in the RGB space. Points with the same color come from the same movement. \textit{S}, \textit{T}, \textit{ST}, \textit{FT}, \textit{SR}, \textit{MR} represent six movements in PURE dataset.}
	\label{fig:pointLayout}
\end{figure}

Next, we investigate the spatial distribution in the RGB space of the trace to answer the second question. We project the detrended traces into the RGB space. All the points in the RGB space are divided into several segments. The segment refers to a point cloud, which is extracted from the video coming from all subjects under one category (\emph{e.g.}, \textit{S}, \textit{T}, \textit{ST}, \textit{FT}, \textit{SR}, and \textit{MR}). We plot the distribution of the points in Fig. \ref{fig:pointLayout}. From the results, we can see that all categories have nearly the same spatial distribution. In addition, we observe that the periodic vibration trajectory of the detrended trace is in 3D space, not just the movement in the single direction of the R, G, B axes. The vibration in each dimension is a decomposed motion of the trajectory. The motion trajectory in the RGB space is the result of the simultaneous action of motion-induced variations and pulse-induced subtle color changes. The motion in PURE is also periodic, as typically occurs in \textit{ST}, \textit{FT}, \textit{SR}, and \textit{MR}. POS and ICA take into account the spatial layout of raw data when extracting the HR, so they are more robust than the proposed method under movement in normal light conditions, such as \textit{T}, \textit{FT}, \textit{SR}, and \textit{MR}. The proposed method performs equally well in the case of \textit{S} and \textit{ST}. For camera quantization noise, is characterized by intensity and color fluctuations above and below the actual image intensity, and has the same effect on the RGB three channels. However, we know that physiological signal is well preserved in the green channel and that the quality is poor in the other two channels. In this case, POS and ICA will introduce more quantization noise, so that it affects the accuracy of HR extraction.

\begin{figure}[!htb]
	\begin{center}
		\includegraphics[width=0.6\linewidth]{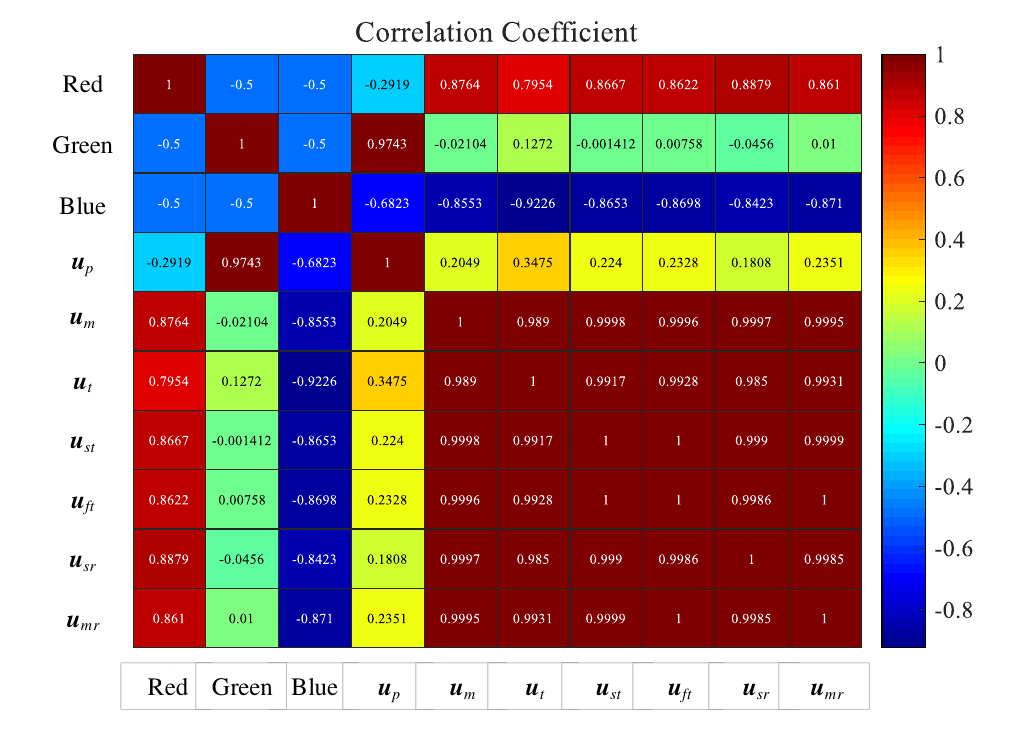}
	\end{center}
	\caption{Correlation analysis among pulse, motions, and coordinate system. The correlation matrix is shown as a heatmap where the intensities are in the range [-1, 1], and the color scheme looks like the color bar on the right. The entries with higher energy (red) represent a higher similarity between traces and vice versa.}
	\label{fig:corr}
\end{figure}

To explore the difference between quantization noise and artifacts caused by motion, we evaluate the distribution of the pulse and motion vectors learned from PURE dataset. Let $\bm{u}_{k}$ be the unit vector learned from the $k$-th category of the PURE dataset, where $k \in \left \{p, m, t, st, ft, sr, mr \right \}$. $\bm{u}_{p}$ and $\bm{u}_{m}$ denote the final pulse and motion vectors, respectively. $t, st, ft, sr, mr$ represent the motion vectors in the case of \textit{T}, \textit{ST}, \textit{FT}, \textit{SR}, and \textit{MR}. Based on the video data from the PURE dataset, we learn unit vectors $\bm{u}_{k}$ according to the method applied in \cite{Zhao_BSPC}. The results are reported in Table \ref{tab:vector}. After that, we measure the correlation coefficients between pulse and motion vectors, as shown in Fig. \ref{fig:corr}. From the test results in Fig. \ref{fig:corr}, it can be seen that the pulse signal is closely related to the green channel, whereas the motion artifact is linked to the R channel. Especially, camera quantization noise is independent of each other on the three axes of the RGB space.

\begin{table}[]
	\caption{The vector distribution of pulse and motions in RGB space. The columns of the vector represent the value in the three directions of the red, green, and blue axis.}
	\begin{center}
			\begin{tabular}{@{}cccccccc@{}}
				\toprule
				\textbf{Category} & $\bm{u}_{p}$ & $\bm{u}_{m}$
				& $\bm{u}_{t}$ & $\bm{u}_{st}$ & $\bm{u}_{ft}$
				& $\bm{u}_{sr}$ & $\bm{u}_{mr}$ \\ \midrule
				\textbf{Value} & $\begin{bmatrix} 0.4806\\ 0.7874\\ 0.3860 \end{bmatrix}$
				& $\begin{bmatrix} 0.7870 \\ 0.5368 \\ 0.3042 \end{bmatrix}$
				& $\begin{bmatrix} 0.7462 \\ 0.5818\\ 0.3235 \end{bmatrix}$
				& $\begin{bmatrix} 0.8010 \\ 0.5346 \\ 0.2695 \end{bmatrix}$
				& $\begin{bmatrix} 0.7925 \\ 0.5407 \\ 0.2822 \end{bmatrix}$
				& $\begin{bmatrix} 0.7687 \\ 0.5394 \\ 0.3437 \end{bmatrix}$
				& $\begin{bmatrix} 0.7454 \\ 0.5582 \\ 0.3644 \end{bmatrix}$ \\ \bottomrule
			\end{tabular}
	\end{center}
	\label{tab:vector}
\end{table}

\subsection{Influence of camera parameters}

The amplitude of pulse-introduced variation is subtle and the presence of noise is associated with the camera parameters, such as video compression, frame rate, and resolution.

\textbf{Video compression.} For lossy compression, videos usually suffer from data loss after compression. The rPPG measurement also suffers from this issue \cite{Zhao_Com}. For the MIHR dataset which is not compressed, the proposed method performs better than the existing methods. Similarly, our method performs the robustness and generalization on the PURE dataset, which is compressed into PNG format, but the pulse signals are well preserved. It is demonstrated that the performance of the proposed method is stable as long as the physiological signal is not destroyed by compression.

\textbf{Frame rate, and resolution.} The MIHR and PURE datasets were both captured with an RGB camera at a frame rate of 30 Hz with a cropped resolution of $640\times480$ pixels. During signal acquisition, the sampling rate is 30 Hz, ensuring sufficient signal sampling points. In this case, the resolution of the video has little effect on the algorithm. 

\section{Conclusions}
\label{sec:CON}
The pulsatile signal preserved in the video can be damaged by insufficient illuminance, thereby causing the failure of remote HR estimation. We performed an analysis of the digital camera quantization noise and its influence on remote HR estimation and obtained its characteristics in comparison with the pulsatile signal in low-light conditions. We estimated the SNR and precision of state-of-the-art rPPG algorithms to verify their performance and robustness on a wide variety of light illuminance. Our results suggest that the previous methods fail to deal with low illuminance. We propose a solution based on the characteristics of digital camera quantization noise for remote HR estimation under low-light conditions. The analysis reveals that low-light enhancement on video frames cannot amplify the pulsatile signal. Thus, the components associated with camera artifacts are considered and removed from raw trace to obtain a clean rPPG signal. The proposed method is evaluated on a public, large-scale dataset MIHR with various illumination intensities and a publicly available dataset PURE which is recorded under normal light conditions. The results demonstrate that our method outperforms state-of-the-art methods in conditions where the illumination is insufficient, and achieved competing results under normal light conditions.

\section*{Conflict of interest}
Authors declare that they have no conflict of interest.

\section*{Funding}
This work was supported by the National Nature Science Foundation (NSFC) of China under Grant No.s 61903336, 61620106012 and the Zhejiang Provincial Natural Science Foundation under Grant No. LY21F030015.

\section*{Ethical Approval}
The ethical approval is not required for this study.

\section*{Dataset}
\href{https://xilin1991.github.io/Large-scale-Multi-illumination-HR-Database/}{Large-scale Multi-illumination HR (MIHR) Dataset}

\href{https://www.tu-ilmenau.de/neurob/data-sets-code/pulse-rate-detection-dataset-pure}{PUlse RatE Detection (PURE) Dataset}

\bibliographystyle{unsrtnat}
\bibliography{mybib}  






\end{document}